\documentstyle[aps,epsf,twocolumn]{revtex}
\newcommand{\be}{\begin{eqnarray}}
\newcommand{\ee}{\end{eqnarray}}
\begin{document}

\draft
\title{\bf  Dilepton and Photon Emission Rates from a Hadronic Gas}

\author{{\bf James V. Steele}$^1$,
{\bf Hidenaga Yamagishi}$^2$  and {\bf Ismail Zahed}$^1$}

\address{$^1$Department of Physics, SUNY, Stony Brook, New York 11794, USA;\\
$^2$4 Chome 11-16-502, Shimomeguro, Meguro, Tokyo, Japan. 153.}
\date{\today}
\maketitle

\begin{abstract}
We analyze the dilepton and photon emission rates from a hadronic gas
using chiral reduction formulas and a virial expansion.  The emission
rates are reduced to pertinent vacuum correlation functions, most of
which can be assessed from experiment. Our results indicate that in
the low mass region, the dilepton and photon rates are enhanced
compared to most of the calculations using chiral Lagrangians.  The
enhancement is further increased through a finite pion chemical
potential.  An estimate of the emission rates is also made using
Haag's expansion for the electromagnetic current.  The relevance of
these results to dilepton and photon emission rates in heavy-ion
collisions is discussed.
\end{abstract}
\pacs{}
\narrowtext

{\bf 1. \,\,\,} Recent fixed target experiments at the CERN SPS using
ultra-relativistic heavy-ion collisions have reported an excess of
dielectrons (CERES) \cite{CERES} and dimuons (HELIOS3) \cite{HELIOS}
over a broad range of dilepton invariant mass starting from the
two-pion threshold.  Some enhancement in the photon emission rate has
been reported by WA80 \cite{WA80} although the direct photon
measurements by CERES and HELIOS3 seem to suggest that the excess is
within statistical errors when the Dalitz decays are subtracted
out. The latter constitute 90\% of the photon spectrum below the
two-pion threshold.

Enhancements in either the dilepton or photon spectra do not seem to
follow from calculations using chiral Lagrangians in hadronic phase
\cite{KAPUSTA,GAS} or high temperature QCD for the quark-gluon phase
\cite{PLASMA}, although neither of them is assumption free for
temperatures around the pion mass. A rate departure from p-A
collisions may be indicative of some medium modifications \cite{BROWN}
in the hadronic phase, although the rate may be enhanced or
depleted \cite{UPDO} depending on model assumptions.

In view of this, it is important that the rate calculations are
reassessed in as model independent way as possible. Also, since the
dilepton and photon emission rates originate from the same
current-current correlation $albeit$ with different kinematics, it is
important that both rates are assessed simultaneously in a consistent
manner. Better photon and dilepton measurements are yet to come,
putting to test our understanding of the strong interactions in the
hot hadronic phase.

In this letter we will show that in a thermal hadronic environment
with a temperature near or below the pion mass the dilepton and photon
emission rates are directly related to the absorptive parts of the
scattering amplitudes of virtual and real photons off an arbitrary
number of on-shell pions.  These scattering amplitudes are reducible
to pertinent vacuum correlation functions using chiral reduction
formulas \cite{MASTER}.  To leading order in the pion density, the
emission rates are constrained by data from electro-production and
-annihilation, tau-decays, pion radiative decay and two-photon fusion
reactions. Our results indicate that in the low mass region the
dilepton and photon rates are enhanced in comparison with calculations
using explicit reaction processes from chiral Lagrangians.  There is
also an enhancement for finite pion chemical potential in both
cases. For comparison, we also give a direct estimate for the emission
rates using Haag's expansion for the electromagnetic current. The
relevance of our results to lepton and photon emission from heavy-ion
collisions is discussed.

\vskip .5cm
{\bf 2.\,\,\,}
In a hadronic gas in thermal equilibrium, the rate ${\bf 
R}$ of dileptons produced in a unit four volume follows from the thermal 
expectation value of the electromagnetic current-current correlation 
function \cite{LARRY}. For massless leptons with momenta $p_1, p_2$,
the rate per unit dilepton momentum $q =p_1+p_2$ is given by 
\be
\frac {d{\bf R}}{d^4q} = -\frac{\alpha^2}{6\pi^3 q^2}\,\,
\,\,{\bf W} (q)
\label{1}
\ee
where $\alpha =e^2/4\pi$ is the fine structure constant,
\be
{\bf W} (q) = \int\!\! d^4x \, e^{-iq\cdot x} \,
{\rm Tr} \left(e^{-({\bf H}-F)/T} \,\,{\bf J}^{\mu} (x){\bf J}_{\mu}
(0)\right), 
\label{2}
\ee
$e{\bf J}_{\mu}$ is the hadronic part of the electromagnetic current,
${\bf H}$ is the hadronic Hamiltonian, $F$ the free energy, $T$ the
temperature, and the trace is over a complete set of hadron states. We
have set the pion chemical potential to zero. The results for a non-zero
pion chemical potential are discussed in section 5.  For leptons with mass
$m_l$, the right-hand side of (\ref{1}) is multiplied by
\be
(1+\frac {2m_l^2}{q^2})(1-\frac{4m_l^2}{q^2})^{\frac 12}.
\label{2.1}
\ee

{}From the spectral representation and symmetry, the rate may also be
expressed in terms of the absorptive part of the time-ordered
function 
\be
{\bf W} (q) = \frac 2{1+e^{q^0/T}} \,\,{\rm Im}{\bf
W}^F(q)   
\label{3}
\ee
\be
{\bf W}^F (q) = i\int d^4x \, e^{iq\cdot x} \,
{\rm Tr} \left(e^{-({\bf H}-F)/T} \,\,T^*{\bf J}^{\mu} (x){\bf
J}_{\mu} (0)\right). 
\nonumber
\ee

{}From here on we take $T \leq m_{\pi}$ and consider only pion
states. Expanding the trace in terms of incoming pions, and summing
over disconnected pieces leads to the virial expansion
\be
{\bf W}^F (q) = &&i \int d^4x \, e^{iq\cdot x} \,
<0| \,T^* {\bf J}^{\mu}(x) {\bf J}_{\mu} (0)|0> \nonumber\\
&&+
\sum_a  i \int\,\frac {d^3k}{(2\pi)^3} \frac {n(\omega_k )}{2\omega_k} 
\int d^4x\, e^{iq\cdot x}\nonumber\\&&\times
<\pi^a(k)\, {\rm in}| \,T^* {\bf J}^{\mu}(x) {\bf J}_{\mu}(0) 
|\pi^a (k)\, {\rm in}>_{\rm conn.}
\nonumber\\
&&+\frac 1{2!} \sum_{a,b}  \int\,\frac {d^3k_1}{(2\pi)^3}
\frac {d^3k_2}{(2\pi)^3} \frac {n(\omega_{k_1} )}{2\omega_{k_1}}
\frac {n(\omega_{k_2} )}{2\omega_{k_2}} 
\nonumber\\ &&\times i\int d^4x\, e^{iq\cdot x}
<\pi^a(k_1)\pi^b (k_2)\, {\rm in}|\nonumber\\&&\times
 \,T^* {\bf J}^{\mu}(x) {\bf J}_{\mu}(0) 
|\pi^a (k_1)\pi^b (k_2)\, {\rm in}>_{\rm conn.}\nonumber\\
&&+ ...
\label{4}
\ee
with the pion energy $\omega_k=\sqrt{m_{\pi}^2 +k^2}$, and the thermal
pion distribution $n(\omega ) = 1/(e^{\omega/T} -1)$.  The matrix
elements in (\ref{4}) correspond to the forward scattering amplitudes
of a virtual photon off on-shell pions, $\gamma^*\, n\pi\rightarrow
\gamma^* \,n\pi$. 

To leading order in the pion density,
\be
n = 3 \int \frac{d^3k}{(2\pi)^3} \,\,n(\omega_k),
\nonumber
\ee
only the first two terms in (\ref{4}) contribute. A slight refinement
is to note that the reduction of one pion is associated with a factor
of $1/f_{\pi}$, where $f_{\pi} =93$ MeV is the pion decay constant, so
that the dimensionless expansion parameter should be $\kappa =
n/2m_{\pi} f_{\pi}^2$. According to Ref. \cite{GOITY}, a pion gas with
a temperature $T \leq m_{\pi}$ is characterized by a pion mean free
path of the order of 2-3 fm over a wide range of momenta, with a pion
mean distance of the order of 2 fm. This gives $\kappa = 0.3$, so the
truncation should be reasonable.

The absorptive part of the vacuum contribution is directly related to
data through $e^+e^-$ annihilation into hadrons \cite{HUANG}. The
forward scattering amplitude $\gamma^*\pi\rightarrow\gamma^*\pi$ is
unfortunately not measurable.  However, it can be constrained by data
as we now discuss.

\vskip .5cm
{\bf 3.\,\,\,}
Decomposing the electromagnetic current into the isovector part ${\bf
V}^3$ and the isoscalar part ${\bf B}$, the forward
$\gamma^*\pi\rightarrow \gamma^*\pi$ matrix element involves three
types of time-ordered correlators ${\bf BB}$, ${\bf BV}$, and ${\bf
VV}$. In pionic states, the ${\bf BB}$ and ${\bf BV}$ correlators are
expected to be small. Indeed, in the soft pion limit, conventional
PCAC gives zero for both matrix elements. The former because ${\bf B}$
commutes with the axial charge, and the latter because the vacuum is
isospin invariant. The ${\bf VV}$ correlator in the one-pion state
${\bf W}^F$ can be reduced to pertinent vacuum correlators using
chiral reduction formulas \cite{MASTER}. The result is ($\nu=k\cdot
q$)
\be
&&{\rm Im} \,{\bf W}^F (q)= -3q^2 {\rm Im} \;{\bf \Pi}_V(q^2)
\nonumber\\
&&{}+\frac1{f_\pi^2}\int\!\!\frac{d^3k}{(2\pi)^3} 
\frac{n(\omega_k)}{2\omega_k} {\bf W}^F_1 (q,k) + {\cal
O}\left(\kappa^2 \right)
\label{5}
\ee
with 
\be
&&{\bf W}^F_1(q,k) = 12q^2 {\rm Im}\; {\bf \Pi}_V(q^2)
\nonumber\\
&&{}-6(k+q)^2 {\rm Im}\; {\bf \Pi}_A\left( (k+q)^2 \right) +
(q\rightarrow -q)
\nonumber\\
&&{}+8(\nu^2-m_\pi^2 q^2) {\rm Im}\;
{\bf \Pi}_V(q^2) 
\nonumber\\
&&\;\;\;\times{\rm Re} \left( \Delta_R(k+q) + \Delta_R(k-q) \right)  
\nonumber\\
&&{}+3m_\pi^2 f_\pi \int\!\! d^4x d^4y \; e^{iq\cdot(x-y)}
\nonumber\\
&&\;\;\;\times
{\rm Im}\; \langle 0 | T^* {\bf V}_\mu^3(x) {\bf V}^{\mu,3} (y)
\hat{\sigma}(0) | 0 \rangle 
\nonumber\\
&&{}-k^\alpha k^\beta \int\!\! d^4x d^4y d^4z \; e^{iq\cdot(x-y)}
e^{-ik\cdot z}
\nonumber\\
&&\;\;\;\times {\rm Im}\; i\langle 0 | T^* {\bf
V}^3_\mu(x) {\bf V}^{\mu,3}(y) {\bf j}^a_{A\alpha}(z) {\bf
j}^a_{A\beta}(0) | 0 \rangle
\nonumber\\
&&{}+k^\beta \epsilon^{a3e} {\rm Im}\;
i\left(\delta^\alpha_\mu - (2k+q)_\mu (k+q)^\alpha \Delta_R(k+q) \right) 
\nonumber\\
&&\;\;\;\times \int\!\! d^4x d^4y \; e^{ik\cdot(y-x)} e^{-iq\cdot x}
\langle 0 | T^* {\bf j}^e_{A\alpha}(x) {\bf
j}^a_{A\beta}(y) {\bf V}^{\mu,3}(0) | 0 \rangle
\nonumber\\
&&{}+(q \rightarrow -q) + (k \rightarrow -k) + (q, k \rightarrow -q, -k)
\label{6}
\ee

where $\Delta_R (k)$ is the retarded pion propagator 
\be
\Delta_R (k) = {\bf PP} \frac 1{k^2-m_{\pi}^2} -i\pi {\rm sgn } (k^0 )
\delta (k^2 -m_{\pi}^2 ),
\label{7}
\ee 
${\bf j}_A$ is the axial vector current with the one-pion contribution
subtracted, ${\bf V}$ is the isovector current discussed above, and
$\hat {\sigma}$ is a scalar density related to $\overline{q}q$
\cite{MASTER}.  ${\bf \Pi}_A$ and ${\bf \Pi}_V$ are the transverse
part of the axial $<0|T^*{\bf j}_A {\bf j}_A |0>$ and vector
$<0|T^*{\bf VV}|0>$ correlators respectively. The absorptive part of
${\bf \Pi}_V$ follows from electroproduction data, while the
absorptive part of ${\bf \Pi}_A$ follows from tau decay data.

Note that the longitudinal part of the axial correlator does not
contribute here.  Taking the chiral limit ($m_\pi=0$) and for a
zero-momentum pion,
\be
{\bf W}^F_1(q,0) = 
12q^2 {\rm Im} \left( {\bf \Pi}_V(q^2) - {\bf \Pi}_A(q^2) \right)
\nonumber
\ee
which reproduces the results obtained in \cite{HUANG,TPCAC}. As
already indicated, eqs. (\ref{5}-\ref{6}) only takes care of the
isovector part of the electromagnetic current, both at zero and first
order in the virial expansion. The inclusion of the isoscalar
contribution to the zeroth order part leaves our arguments unchanged
(at the 1\% level) except around the phi region.

The third term of (\ref{6}) contains the principle value integral from
(\ref{7}). However, there is no pole within the range of integration
unless $q^2=2m_\pi q_0$. This term is proportional to $q^2{\rm Im}
{\bf \Pi}_V (q^2)$ vanishing at $q^2=0$.

The last three terms of (\ref{6}) are vacuum correlators which appear
in other low-energy processes and are also constrained by data. The
${\bf j}_A {\bf j}_A {\bf V}$ correlator appears in the radiative
decay of the pion \cite{MASTER}.  Empirically, the contribution of
this term is small at threshold. There are two ways in which this can
be checked analytically: a resonance saturation \cite{SERGEI} and a
one-loop expansion \cite{MASTER}. Both of these methods give zero for
this correlator.

The ${\bf j}_A{\bf j}_A{\bf V} {\bf V}$ and ${\bf V V}\hat{\sigma}$
correlators appear in the two-photon fusion process \cite{MASTER}.
Resonance saturation for both of these gives on the order of a 1\%
correction and the one-loop result contributes about 1\% in the low
mass region and even less at higher energies. Both descriptions are in
agreement with data from $\gamma\gamma \rightarrow \pi^+\pi^-$
\cite{PLUS} and $\gamma\gamma \rightarrow \pi^0 \pi^0$
\cite{ZERO}. Hence, the last three correlators in (\ref{6}) may be
ignored and the rest can be extracted from data. This information will
be used in section 5 to assess the dilepton and photon rate.

The present analysis can be readily carried out to the photon emission
rate through
\be
q^0 \frac {d{\bf R}}{d^3 q} = -\frac{\alpha}{4\pi^2} 
\,\,{\bf W} (q)
\label{8}
\ee
with $q^2=0$. The constraints of current conservation and chiral
symmetry apply equally well to the photon emission rate. As indicated
above, it is important to assess both emission rates simultaneously
for consistency.  In our case, the only non-zero contribution comes from 
the ${\bf \Pi}_A$ term in eq.~(\ref{6}).

\vskip .5cm
{\bf 4.\,\,\,} For completeness, we mention another expansion scheme
called the Haag expansion. From the pion electromagnetic form factor
\be
&&<\pi^a (p') \pi^b (p) \,{\rm in}| {\bf S} {\bf J}_{\mu} (x) | 0 > =
\nonumber\\
&&=i\epsilon^{a3b} (p'-p)_{\mu} \,\,e^{i(p+p')\cdot x}\,\, {\bf F}_V
((p+p')^2) 
\nonumber
\ee
and the crossed versions, we obtain the Haag expansion
\be
&&{\bf S} {\bf J}_{\mu} (x) = i\epsilon^{a3b} \int \frac{d^3p}{(2\pi)^3}
\frac{d^3p'}{(2\pi)^3} \frac1{2\omega_p} \frac1{2\omega_{p'}} \nonumber\\ 
\Bigg(&& (p+p')_{\mu} \, {\bf F}_V ((p-p')^2)\,\,a_{\rm in}^{a\dagger} (p' ) 
a_{\rm in}^b (p) e^{-i(p-p')\cdot x}\nonumber\\
+&&\frac 12 (p'-p)_{\mu} \, {\bf F}_V ((p+p')^2)\,\,a_{\rm
in}^{a\dagger} (p' )  
a_{\rm in}^{b\dagger} (p) e^{i(p+p')\cdot x}\nonumber\\
+&&\frac 12 (p-p')_{\mu} \, {\bf F}_V ((p+p')^2)\,\,a_{\rm in}^{a} (p' ) 
a_{\rm in}^b (p) e^{-i(p+p')\cdot x} \Bigg)\nonumber\\
+&& {\cal O}(\pi_{\rm in}^4 )
\nonumber
\ee
and similarly for ${\bf J}_{\mu} (y) {\bf S}^{\dagger}$, where ${\bf
S}$ is the  
hadronic S-matrix. Inserting this into (\ref{2}), and 
using energy-momentum conservation along with Wick's theorem, we obtain
\be
{\bf W} (q) = && -\frac 1{8\pi |\vec q |} (q^2-4m_{\pi}^2) 
|{\bf F}_V (q^2) |^2 \nonumber\\
&&\times \theta (q^2-4m_{\pi}^2) \, \int_{\omega_-}^{\omega_+}
d\omega\,\, n(\omega ) \, n( q^0-\omega )
\nonumber
\ee
\be
\omega_{\pm} = \frac 12q^0 \pm \frac 12 |\vec q | 
(1-\frac{4m_{\pi}^2}{q^2})^{\frac 12}.
\label{9}
\ee
The form factor ${\bf F}_V (q^2)$ for $q^2 \geq 4m_{\pi}^2$ is
directly measured from electro-production.  For $q^2 <0$ it follows
from pion scattering on hydrogen targets. In both cases, rho dominance
holds to a good accuracy. This is similar to the first term in the
virial expansion (\ref{4}), but the Haag expansion is presumably less
controlled. In particular, the photon rate is zero in this
approximation.  We note that (\ref{9}) in the $\vec{q}=0$ limit has
already been used in \cite{KAPUSTA}.

\begin{figure}
\begin{center}
\leavevmode
\epsfxsize=3.375in
\epsffile{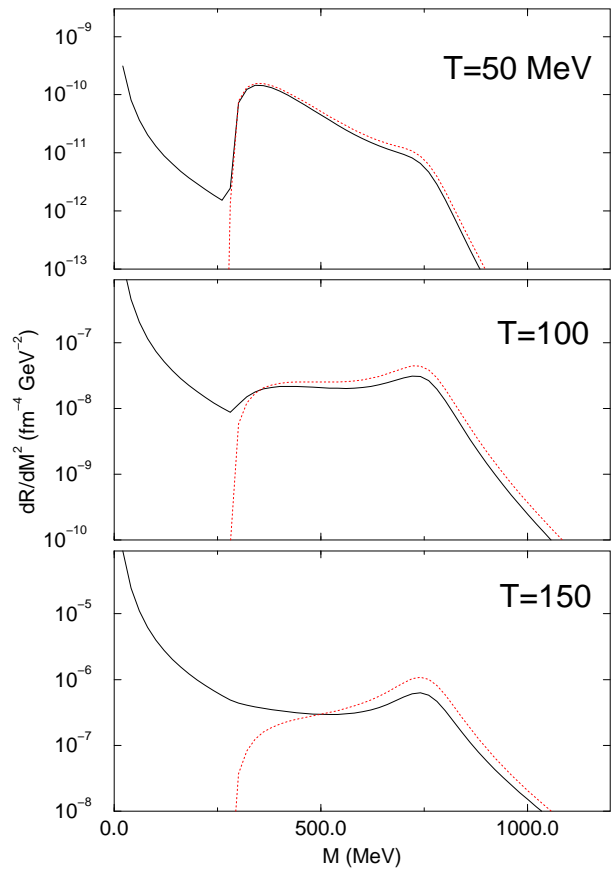}
\end{center}
\caption{The total integrated rate for dielectrons from a hadronic
gas of $T=50, 100$ and $150$ MeV.  The solid line is from the virial
expansion and the dotted line from the Haag expansion.}
\end{figure}

\vskip .5cm
{\bf 5. \,\,\,}
In Fig.~1, we show the numerical results following from the above
analysis for the dielectron rate as a function of the invariant dielectron
mass $M=\sqrt{q^2}$, all the way up to the $a1$ mass for the temperatures
T = 50, 100, 150 MeV.  The change of variables to the rapidity $y$
and magnitude of the transverse momentum $q_\perp$ \cite{LARRY}
\be
\frac{d{\bf R}}{d^4q} = \frac2{\pi} \frac{d{\bf R}}{dM^2\, dy\, dq_\perp^2}
\nonumber
\ee
and an integration over $y$ and $q_\perp^2$ is performed.  The solid
curve is the result following from (\ref{1}) using (\ref{3}) and
(\ref{5}).  The dotted curve is the result following from (\ref{1})
and the Haag expansion (\ref{9}). For comparison, we also show the
integrated dimuon rate in Fig.~2.  As the temperature increases the
rate for dilepton momentum below the rho pole increases with respect
to the Haag expansion. The rate is always cut off at the dilepton
threshold by phase space (\ref{2.1}).

\begin{figure}
\begin{center}
\leavevmode
\epsfxsize=3.375in
\epsffile{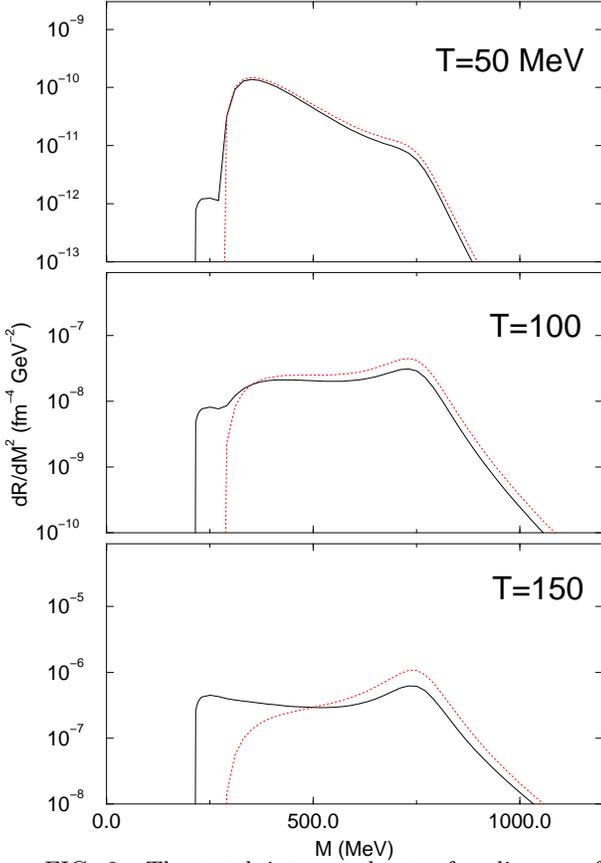}
\end{center}
\caption{The total integrated rate for dimuons from a hadronic
gas of $T=50, 100$ and $150$ MeV.  The solid line is from the virial
expansion and the dotted line from the Haag expansion.}
\end{figure}

For a parameterization of ${\bf F}_V$, we used the common Breit-Wigner
form with a varying width as found in \cite{WEISE} and a small
correction for large $q^2$ to fit the data \cite{STEELE}. 
\be
{\bf F}_V(q^2) = \frac{m_\rho^2+\gamma q^2}{m_\rho^2-q^2-im_\rho
\Gamma_\rho(q^2)} 
\label{10.1}
\ee
with a non-zero width only for $q^2 > 4m_\pi^2$ \cite{WEISE}
\be
\Gamma_\rho(q^2) = 0.149 \frac{m_\rho}{\sqrt{q^2}}
\left( \frac{q^2-4m_\pi^2}{m_\rho^2 -4m_\pi^2} \right)^{3/2} {\rm GeV}
\label{10.2}
\ee
and $\gamma = 0.4$. 

For the terms in (\ref{5}) depending on ${\rm Im} \;{\bf \Pi}_V$, our
results will be sensitive to the momentum dependence of the width. The
KSFR relation implies \cite{MASTER}
\be
{\bf F}_V(q^2) = 1 + \frac{q^2}{2f_\pi^2} {\bf \Pi}_V(q^2) + {\cal
O}\!\left(\frac1{f_\pi^4}\right),  
\nonumber
\ee
so we have fixed ${\bf \Pi}_V$ from ${\bf F}_V$ with the $1/f_{\pi}^4$
terms omitted. Fig.~3 shows this fit against the data taken from the
recent compilation by Huang \cite{HUANG}. Using the fit instead of the
data produces a 10\% uncertainty at the rho pole where the fit is
worst. On the log plot of Fig.~1 this is a negligible difference.

\begin{figure}
\begin{center}
\leavevmode
\epsfysize=6cm
\epsffile{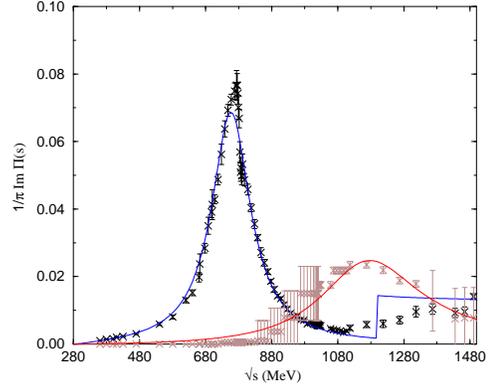}
\end{center}
\caption{Data for the vector and axial spectral densities 
and the fits we used.}
\end{figure}

The axial channel is dominated by the $a1$(1260).  A Breit-Wigner fit
for ${\bf \Pi}_A$ works well with $(f_A,m_A)=(190, 1210)$ MeV through
the substitution $m_{\rho}^2+\gamma q^2\rightarrow f_A^2$,
$m_{\rho}\rightarrow m_A$, and $\Gamma_{\rho}(q^2)\rightarrow
\Gamma_A(q^2)$ in (\ref{10.1}). In (\ref{10.2}), the threshold is
moved to $9m_\pi^2$ with $(m_\rho,4m_\pi^2)\rightarrow (m_A,9m_\pi^2)$
and the overall constant is changed so that $\Gamma_A(m_A^2) = 0.4$
GeV.  The comparison of this fit to the data is also shown in
Fig.~3. We note that fixed-width calculations tend to overestimate the
tail of the $a1$ in the low mass region by a large factor.

In Fig.~4, we show the non-integrated results for back-to-back
($\vec{q}=0$) dielectrons with $T=50, 100, 150$ MeV. As in the fully
integrated rate, our result and the Haag expansion are practically
identical for the lower temperatures and, as the temperature
increases, the region below the two pion threshold increases with
respect to the Haag rate.

The shape of the virial expansion curves in Figs 1, 2, and 4 can be
explained as follows. Above the two-pion threshold, the first term of
eq.~(\ref{5}) dominates giving the rho peak. As the temperature
increases, the ${\rm Im}\;{\bf\Pi}_V$ terms in eq.~(\ref{6}) start to
play a part and, since they give a negative contribution to the rate,
decrease the rho peak by about 10\% for $T=150$ MeV. The ${\rm
Im}\;{\bf \Pi}_V$ terms are proportional to $\Gamma_\rho(q^2)$ and as
a consequence all vanish below the two-pion threshold. This part is
fully consistent with the Haag rate for all temperatures.

The main difference comes from the inclusion of the ${\rm Im}\;{\bf
\Pi}_A$ term. It is the only contribution below the two-pion
threshold. The reason it picks up the $a1$ pole contribution even at low
$q^2$ is because, in contrast to the result in the chiral limit, the
axial spectral density is integrated over all momentum in the thermal
averaging. This weakens the contribution from the $(k+q)^2$ prefactor
in eq.~(\ref{6}) therefore allowing the $1/q^2$ dependence in
(\ref{1}) to dominate at low $q^2$. Integrating the rate pronounces
this even more and is only cut off at the dilepton threshold by
(\ref{2.1}).

\begin{figure}
\begin{center}
\leavevmode
\epsfxsize=3.375in
\epsffile{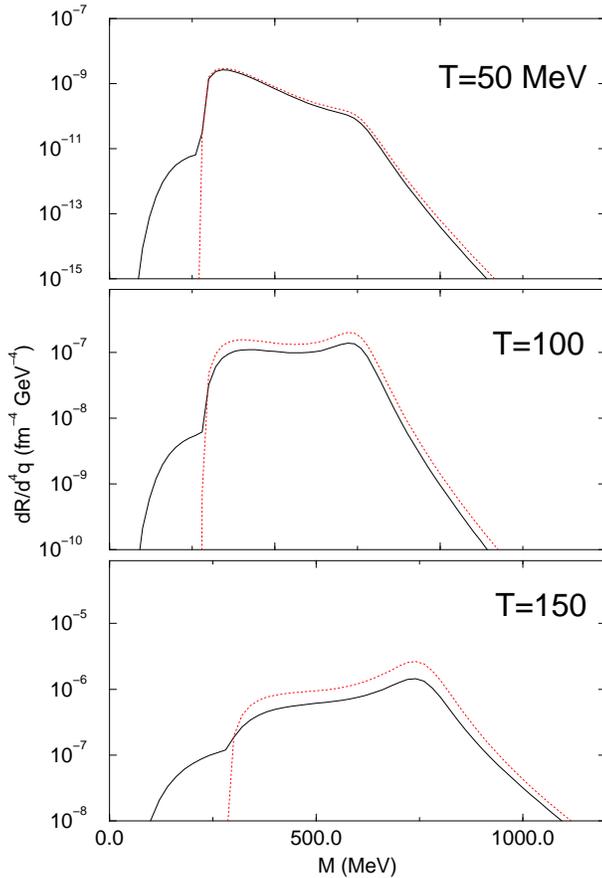}
\end{center}
\caption{The back-to-back rate for dielectrons.  The virial expansion
is the solid line and the Haag expansion is the dotted line.}
\end{figure}

This enhancement can be physically understood as follows. Since
$2\;{\rm Im}\;{\bf W}^F(q) = {\bf W} (q) + {\bf W} (-q)$, the low-mass
region is enhanced by processes such as ${\bf X}\rightarrow \pi
e^+e^-$ through ${\bf W} (-q)$. These are reminiscent of the Dalitz
decays. This description, however, is very qualitative since the
relationship between the physical reactions in the emission rate and
the imaginary part gets blurred by the virial expansion.  In
comparison with most dilepton rate calculations
\cite{KAPUSTA,HUANG,MANY,RUDAZ}, our results show a substantial
enhancement in the low mass dilepton region (2-4 $m_{\pi}$), for
$T=100$-150 MeV.

For dileptons of mass above the $a1$ pole, our emission rate is also
larger than the one expected from the Haag expansion due to the
contribution from the continuum in the rho channel as well as the
effects from the $a1$.  At $T=150$ MeV the enhancement is about an
order of magnitude in the dielectron rate.

If we were to assume an acceptance factor of about $1/100$
(corresponding to the experimental cuts in $q_\perp$ and $y$), a
lifetime for the thermal gas of 5 fm$/c$, and a normalization of $2n$
(due to charged particle multiplicities), our dielectron rate
distribution at a temperature of 100-150 MeV is similar in shape and
magnitude to the physically measured dielectron rate by the CERES
collaboration \cite{CERES}.  In addition, our dimuon rate distribution
for a temperature of 50-100 MeV is similar in shape to those measured
by the HELIOS3 collaboration \cite{HELIOS} prior to the experimental
acceptance cuts. Of course, these are purely heuristic statements that
should motivate a more detailed calculation using a transport
equation.

\begin{figure}
\begin{center}
\leavevmode
\epsfxsize=3.25in
\epsffile{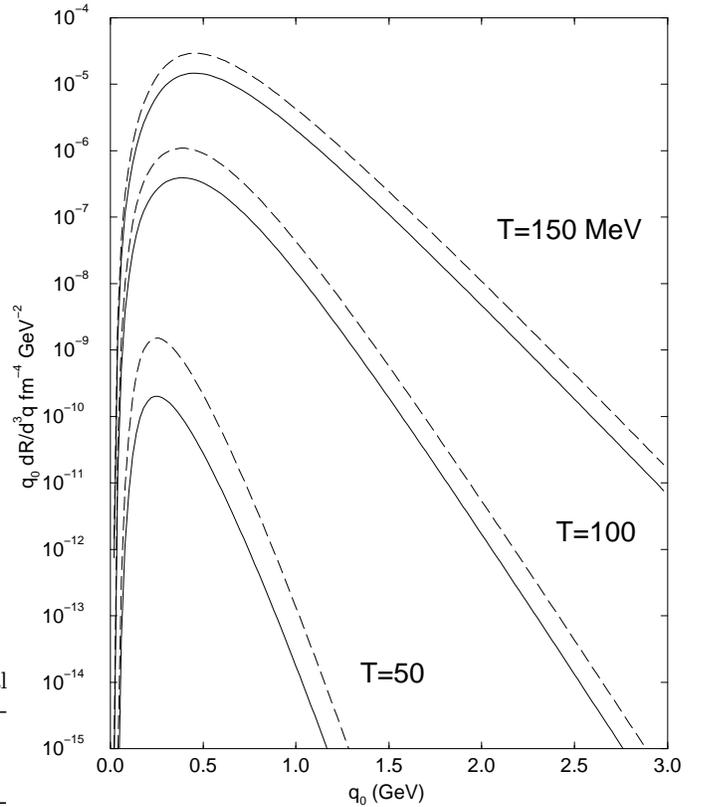}
\end{center}
\caption{The photon emission rate from a hot hadronic gas. The solid
and dotted lines were calculated with a pion chemical potential of 0
and 100 MeV respectively.}
\end{figure}

Fig.~5 shows the photon emission rate versus the photon energy for
$T=50, 100, 150$ MeV. The solid line is the result following from the
virial expansion. As we have noted, there is no contribution from the
Haag expansion to the order quoted.  We note that the the bulk of the
low mass dileptons stem primarily from the photon source.  Our results
are an order of magnitude above those obtained using two-body
reactions with thermal $\rho$'s and $a1$'s \cite{RUDAZ,ALL} at the
peak of the curve. However, above the $a1$ mass, our results are in
agreement with others. We note that the use of a constant width for
$\Gamma_A$ increases the rate by almost an order of magnitude for all
temperatures.

Other calculations \cite{ALL} tend to underestimate the pion electric
polarizability by almost an order of magnitude which could point to
the source of their result being smaller.  In our case, the result is
constrained by the data for ${\bf\Pi}_A$.

The role of the pion chemical potential $\mu_{\pi}$ may be assessed
through the substitution $n(\omega ) = (e^{\omega/T}
-1)^{-1}\rightarrow (e^{(\omega -\mu_{\pi})/T} -1)^{-1}$ in the above
formulas.  A non-zero chemical potential influences all terms which
involve pions.  This includes the entire Haag rate. In contrast, the
first term in the virial expansion, which is its dominant contribution
around the rho pole, is not influenced.

Fig~6. shows the rates for $\mu_\pi=100$ MeV which describes a dilute
system of pions with a mean pion separation of about 2 fm at
$T=100$-150 MeV\cite{GOITY}.  The dilepton integrated rate following
from the virial expansion is enhanced in the low mass region by about
a factor of five. There is no change near the rho pole and above 1 GeV
it is again slightly enhanced. The entire Haag rate, on the other
hand, is increased by about a factor of 10. This is a major
qualitative difference in the two analyses.  This further enhancement
points at the importance of pions off-equilibrium, thereby motivating
a more quantitative analysis in the context of kinetic theory.  The
photon rates are systematically enhanced as shown by the dotted lines
in Fig.~5. In the temperature range $T=100$-150 MeV, the enhancement
is about a factor of two throughout the photon energy spectrum.

\vskip .5cm
{\bf 6.\,\,\,} 
To summarize, we have argued that for a hadronic gas at temperatures
$T \leq m_{\pi}$ the dilepton and photon emission rates were
constrained by data from electroproduction, tau decays, radiative pion
decay and two-photon fusion reactions. Through these constraints, the
rates were assessed using data for the various vacuum correlation
functions and carried out to the first two terms in the virial
expansion (\ref{4}). For completeness, we have compared the results
following from the Haag expansion.

Through chiral reduction formulas, our construction shows the
interdependence between the dilepton and photon rates and other low
energy processes.  We note that the chiral reduction formulas follow
solely from broken chiral symmetry with explicit (2,2) breaking.  They
are a direct generalization of current algebra results from threshold
to general momenta.  In this respect, our analysis is different from
finite temperature arguments using PCAC in the chiral limit
\cite{HUANG,TPCAC}. Our analysis was confined to the isovector part of
the electromagnetic current, and could be improved at higher mass (phi
region) by including the isoscalar contribution.

\begin{figure}
\begin{center}
\leavevmode
\epsfxsize=3.25in
\epsffile{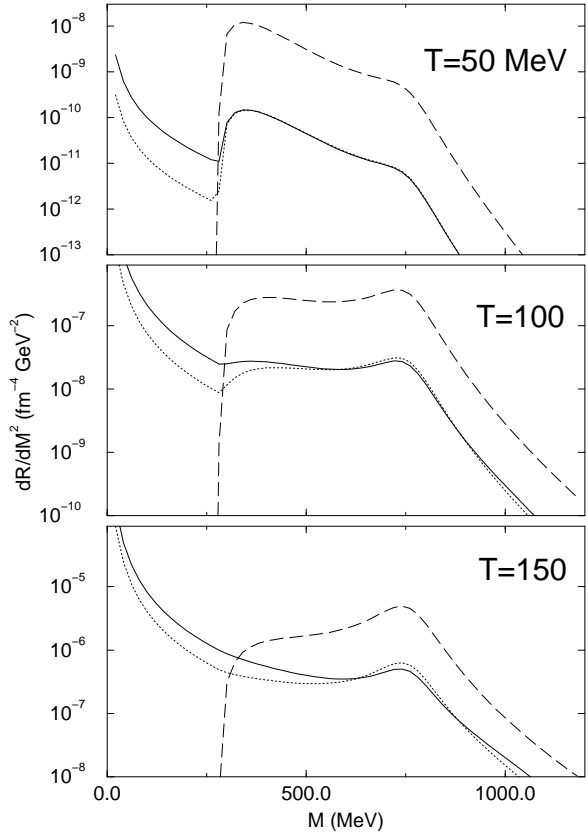}
\end{center}
\caption{The dielectron integrated rates for $T=50, 100, 150$ MeV and
a finite pion chemical potential $\mu_\pi=100$ MeV. The solid line is
from the virial expansion.  The dashed line is from the Haag
expansion.  For comparison, the virial expansion result at zero
chemical potential is shown as the dotted line.}
\end{figure}

Since dilepton or photon emission from heavy-ion collisions involve a
dynamical integration over the real-time history of the hadronic gas
that is expected to expand and cool, our results are only
suggestive. However, they already show a thermal pion gas in
equilibrium yields dilepton and photon rates that are enhanced in the
low mass region compared to most of the calculations using specific
reaction processes.  The inclusion of a pion chemical potential
enhances the entire photon rate and the low and high mass
dilepton rates without affecting the rate distribution around the rho
mass. These results are relevant for present dilepton measurements at
CERES and HELIOS3 as well as future dilepton and photon measurements
at RHIC and LHC.

\vskip 0.6cm
{\bf \noindent  Acknowledgements \hfil}
\vglue 0.4cm
We would like to thank Gerry Brown, Robert Pisarski, Madappa Prakash
and Heinz Sorge for discussions. We are grateful to Zheng Huang for
providing us with his newly compiled data. This work was supported in
part by the US DOE grant DE-FG-88ER40388.

\vskip 1cm
\setlength{\baselineskip}{15pt}

\end{document}